\documentclass[usenatbib]{iaus}
\usepackage{graphicx}

\title[Binary interactions and UV photometry on photo-$z$]
{Binary interactions and UV photometry on photometric redshift}

\author[F. Zhang, L. Li \& Z. Han]
{F. Zhang$^1$, L. Li$^1$ \and Z. Han$^1$}

\affiliation{$^1$National Astronomical Observatories/Yunnan
Observatory, Chinese Academy of Sciences, PO Box 110, Kunming,
Yunnan Province 650011, China \break email:
zhangfh@ynao.ac.cn}

\pubyear{2009}
\volume{262}
\pagerange{000--111}
\date{??? and in revised form ???}
\jname{Stellar Populations: Planning for the Next Decade}
\editors{S.Charlot, ............. \& C. Chiosi, eds.}
\begin{document}

\maketitle

\begin{abstract}
Using the \textit{Hyperz} code (Bolzonella et al. 2000) we present photometric redshift estimates for a random sample of galaxies selected from the SDSS/DR7 and GALEX/DR4, for which spectroscopic redshifts are also available.

We confirm that the inclusion of ultraviolet photometry improves the accuracy of photo-$z$s for those galaxies with $g^\star-r^\star \le 0.7$ and $z_{\rm spec} \le 0.2$.
We also address the problem of how binary interactions can affect photo-$z$ estimates, and find that their effect is negligible.

\end{abstract}

{\bf 1. Introduction}
Redshift is one of the key ingredients in the studies of galaxy formation and evolution. Because that photometry is $\sim$2 orders of magnitude less time-consuming than spectroscopy for a given telescope size, upcoming redshift surveys will exclusively reply on the photometry.

To date three distinct methods have been used to derive photo-$z$: template fitting, empirical train-set and instance-based learning approaches. In the previous works, which the photo-$z$s of galaxies were obtained by using template fitting method, the template spectra came from either single stellar population (SSP) synthesis models or real objects, never from the population synthesis models considering binary interactions.
As for binary stellar population (BSP) synthesis models, Yunnan group (Zhang et al. 2004b, 2005) have considered various of binary interactions in population synthesis models, and drawn the conclusion that the inclusion of binary interactions can affect the overall shape of the spectral energy distribution (SED) of population, in particular, the SED of population in the ultraviolet (UV) passbands is bluer by 2-3 mag at $t \sim 1$\,Gyr if binary interactions are accounted for.

{\bf 2. Method}
The photo-$z$s of galaxies are computed through the \textit{Hyperz} code (Bolzonella et al. 2000), which is a template fitting procedure and adopts a standard $\chi^2$ minimization algorithm.

The template spectral library, required by the \textit{Hyperz} code, consists of 8 theoretical spectral families and 4 mean observed spectra for local E-, Sbc-, Scd- and Irr-type galaxies (Coleman, Wu \& Weedman 1990). The 8 spectral families are for Burst, E, S0-Sd and Irr types of galaxies, respectively, and built with Yunnan SSP or BSP models (zhang et al. 2002, 2004a,b 2005) by using the BC03 software package.

{\bf 3. Sample}
We select 6531 galaxies randomly from SDSS/DR7 and GALEX/DR4 (the matching radius is 6 arcsec), for which the SDSS spectroscopic redshifts $z_{\rm spec}$ are available. We use the SDSS $ugriz$ \textit{modelMag}.

\begin{figure*}
\centering{
\includegraphics[bb=192 200 457 558,height=3.0cm,width=2.4cm,clip,angle=-90]{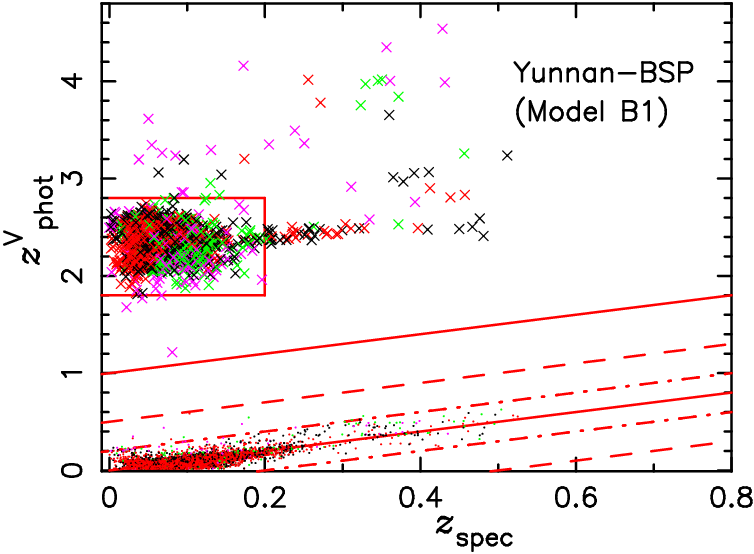}
\includegraphics[bb=192 200 457 558,height=3.0cm,width=2.4cm,clip,angle=-90]{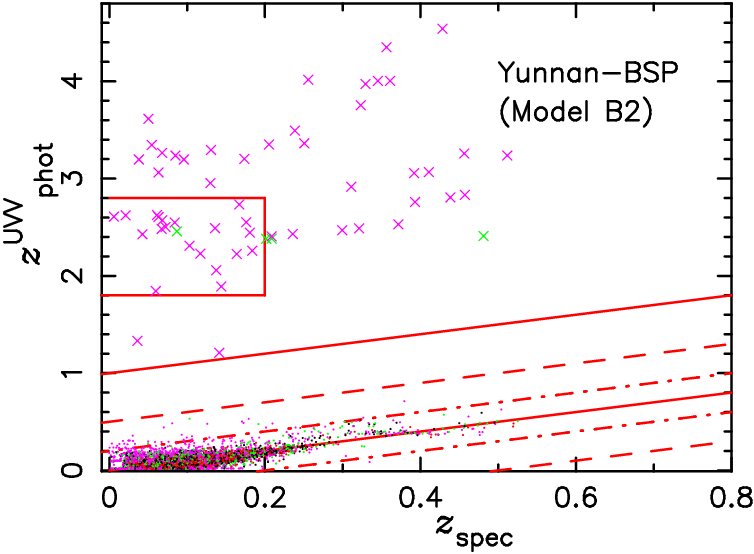}
\includegraphics[bb=192 200 457 558,height=3.0cm,width=2.4cm,clip,angle=-90]{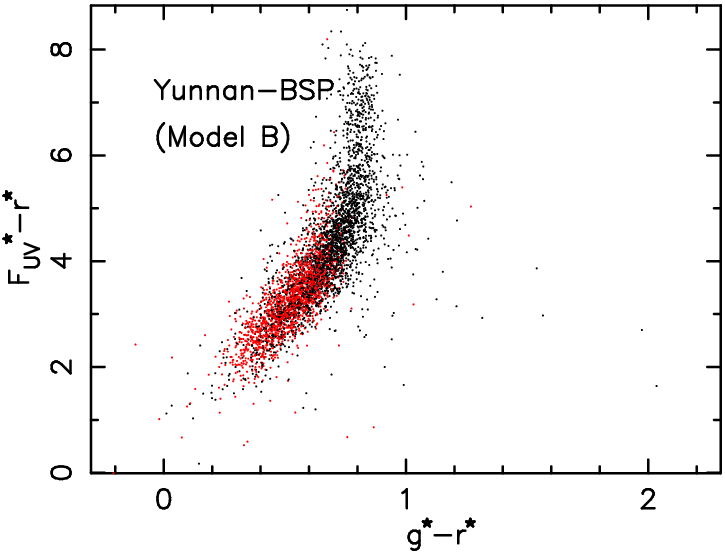}
\includegraphics[bb=192 200 464 558,height=3.0cm,width=2.4cm,clip,angle=-90]{px2-ynbb.ps}
}
\caption{From left to right the 1st and 2nd panels are for the comparison between $z_{\rm spec}$ and $z_{\rm phot}$ of galaxy for models B1 and B2. Black, red, green and purple symbols represent the probability
 $99 \ge P(\chi^2) > 90$, $90 \ge P(\chi^2) \ge 68$ and $P(\chi^2) <68$.
{\bf The 3st panel} is the distribution of galaxies in the $F_{\rm UV}-r$.vs.$g-r$ plane, red symbols are those galaxies of which the accuracy of photo-$z$ has been improved if adding UV photometry. {\bf Right panel} is the probability distribution of photo-$z$ for a galaxy. Black and red lines are for models B1 and B2.}
\label{Fig:lowz-dia-hpl}
\end{figure*}

{\bf 4. Results}
To clarity, we define 4 models: Model A1 uses Yunnan SSP models and $ugriz$ photometry, Model A2 differs from A1 by adding $F_{\rm UV}$ and $N_{\rm UV}$ photometry. Model B1 uses Yunnan BSP models and $ugriz$ photometry, Model B2 differs from B1 by adding $F_{\rm UV}$ and $N_{\rm UV}$ photometry.

{\bf 3.1 UV photometry:}
By comparing the photo-$z$s between models A1 and A2, and those between models B1 and B2, we find $\sim$\,1500 objects with $z_{\rm spec} \le 0.2$ are erroneously identified as high redshift galaxies if only using $ugriz$ photometry,
if including $F_{\rm UV}$ and $N_{\rm UV}$ photometry the number of catastrophic identifications (i.e., $|z_{\rm spec} - z_{\rm phot}| \ge 1.0$) decreases to $\sim$\,120. In the 1st and 2nd panels of Fig.~1 we give the comparisons between $z_{\rm spec}$ and $z_{\rm phot}$ for models B1 and B2.

After all magnitudes are corrected for foreground extinction [using the Schlegel et al. (1998) reddening map and reddening law; Cardelli et al. (1989) reddening law] and \textit{K}-corrected (Blanton \& Roweis 2007), we find those galaxies, which are erroneously identified for models A1 and B1, mainly are bluer galaxies with $g^\star-r^\star \le 0.7$ (see the 3rd panel of Fig.~1).
The reason that including UV photometry can improve the accuracy of photo-$z$s is at $z \le 0.2$ domain no spectral signature can be used in the \textit{Hyperz} procedure, the probability function $P(\chi^2)$ has two peaks because of degeneracy among the fit parameters, if adding UV photometry the degeneration between high and low redshift solutions would disappear (see the right panel of Fig.~1).

{\bf 3.2 Binary interactions:}
By comparing the photo-$z$s of galaxies between models A2 and B2, we find that the effect of binary interactions on photo-$z$ determinations is negligible. The main reason is that significant difference caused by binary interactions is in the $F_{\rm UV}$ passband. while this only occurs at log($\tau) \ge 9$\,yr, leading that binary interactions make $F_{\rm UV}$ significantly bluer only for those systems with shorter star formation timescale ($\tau_{\rm SF} < 1$\,Gyr).
Therefore, we may add some features about the visible/UV flux ratio (such as, $4000{\rm \AA break}/F_{\rm UV}$) in the \textit{Hyperz} package to further constrain the parameters for those galaxies with shorter star formation timescale.

{\bf Acknowledgments} This work was funded by the Chinese Natural
Science Foundation (Grant Nos 10773026, 10673020 \& 10821061) and
by Yunnan Natural Science Foundation (Grant No 2007A113M).

\leftline{\bf References}

\leftline{{Blanton, M., \& Roweis, S.} 2007,
     \textit{AJ}, 133, 734}

\leftline{{Bolzonella, M., Miralles, J.M., \& Pell\,o, R.} 2000,
     \textit{A\&A}, 363, 476}

\leftline{{Coleman, G., Wu, C., \& Weedman, D.} 1980,
     \textit{ApJS}, 43, 393}

\leftline{{Schlegel, D.J., Finkbeiner, D., Douglas, P., \& Davis, M.} 1998,
     \textit{ApJ}, 500, 525}

\leftline{{Zhang, F., Han, Z., Li, L., \& Hurley, J.R.} 2002,
     \textit{MNRAS}, 334, 883}

\leftline{{Zhang, F., Han, Z., Li, L., \& Hurley, J.R.} 2004a,
     \textit{MNRAS}, 350, 710}

\leftline{{Zhang, F., Han, Z., Li, L., \& Hurley, J.R.} 2004b,
     \textit{A\&A}, 415, 117}

\leftline{{Zhang, F., Han, Z., Li, L., \& Hurley, J.R.} 2005,
     \textit{MNRAS}, 357, 1088}









\end{document}